\documentclass[aps,prl,twocolumn,showpacs,10pt,superscriptaddress,preprintnumbers,nofootinbib]{revtex4-1}
\usepackage{epsfig,amssymb,amsmath,psfrag,epstopdf,xcolor}
\pdfoutput=1
\usepackage{graphicx}
\usepackage{subcaption}
\usepackage{ragged2e}
\DeclareCaptionJustification{justified}{\justifying}
 \usepackage[normalem]{ulem}
\usepackage{paralist}
\usepackage{hyperref}
\usepackage{mathrsfs}
\allowdisplaybreaks

\def\beq{\begin{equation}}
\def\eeq{\end{equation}}
\def\bsp#1\esp{\begin{split}#1\end{split}}
\newcommand{\be}{\begin{equation}}
\newcommand{\ee}{\end{equation}}
\newcommand{\bea}{\begin{eqnarray}}
\newcommand{\eea}{\end{eqnarray}}

\def\Fig#1{Fig.~{\ref{#1}}}

\def\cO{{\mathcal O}}

\newcommand{\comment}[1]{}

\newcommand{\td}{\mathrm{d}}

\newcommand{\qqquad}{\qquad\qquad}

\usepackage{slashed}

\begin{document}


\title{Simple Scaling Laws for Energy Correlators in Nuclear Matter}


\author{Carlota Andres}
\affiliation{Center for Theoretical Physics, Massachusetts Institute of Technology, Cambridge, MA 02139, USA}

\author{Fabio Dominguez}
\affiliation{Instituto Galego de F{\'{i}}sica de Altas Enerx{\'{i}}as (IGFAE),  Universidade de Santiago de Compostela, Santiago de Compostela 15782,  Spain}

\author{Jack Holguin}
\affiliation{Consortium for Fundamental Physics, School of Physics \& Astronomy, University of Manchester, Manchester M13 9PL, United Kingdom}

\author{Cyrille Marquet}
\affiliation{CPHT, CNRS, \'Ecole polytechnique,  Institut Polytechnique de Paris, 91120 Palaiseau, France}

\author{Ian Moult}
\affiliation{Department of Physics, Yale University, New Haven, CT 06511}

\preprint{MIT-CTP/5724}


\begin{abstract}
Collider experiments involving nuclei provide a direct means of studying exotic states of nuclear matter.
Recent measurements of energy correlators in both proton-nucleus (p-A) and nucleus-nucleus (A-A) collisions reveal sizable modifications, attributable to nuclear effects, compared to proton-proton (p-p) collisions.
Energy correlators, and their associated light-ray operator product expansion (OPE), allow scaling behaviors of the measured spectrum to be directly mapped to properties of the underlying quantum field theory. 
Here, we demonstrate for the first time how this mapping occurs in nuclear collisions, and highlight how the light-ray OPE characterizes leading nuclear effects.
We show that the leading modification to the energy correlator distribution is characterized by an enhancement of the expectation value of twist-4 light-ray operators, resulting in a scaling for the ratio of the two-point correlator in nuclear matter to that in vacuum  of $\sim 1+a\theta^2$ up to quantum corrections. 
We verify that this leading twist-4 correction accurately describes recent A-A and p-A data, and is thus sufficient to capture the scaling behavior within the angular range measured for jet radii used in nuclear experiments.
Our light-ray OPE based approach lays the groundwork for a rigorous characterization of nuclear modification to energy correlator observables.
\end{abstract}

\maketitle


\emph{Introduction.}---Collider experiments involving large nuclei, such as electron-ion (e-A), proton-ion (p-A), and heavy-ion collisions (A-A), provide valuable insights into the properties of nuclear matter by examining the propagation of energetic partons produced in the underlying hard scattering process \cite{Busza:2018rrf,CMS:2024krd,ALICE:2022wpn,Cao:2020wlm,Apolinario:2022vzg,Cunqueiro:2021wls,Connors:2017ptx,Harris:2024aov,Arratia:2019vju,Brewer:2021kiv,AbdulKhalek:2022hcn,Abir:2023fpo}. Among these hard probes, single-inclusive hadron production is the most theoretically well-understood. In e-A collisions, this process is rigorously described by factorization theorems \cite{Collins:1985ue,Collins:1989gx}, with nuclear modifications encoded in higher-twist operators \cite{Politzer:1980me,Ellis:1982cd,Ellis:1982wd,Jaffe:1983hp,Jaffe:1981td,Jaffe:1982pm,Qiu:1990xy,Qiu:1990xxa} whose matrix elements are enhanced by the size of the nucleus, $A^{1/3}$ \cite{Luo:1993ui,Luo:1994np,Luo:1992eq,Qiu:1991wg,Luo:1992fz,Luo:1991bj,Kastella:1989vd,Kastella:1989ux}.
Under certain assumptions, this framework extends to p-A and A-A collisions \cite{Wang:2001ifa,Guo:2000nz,Wang:2002ri,Osborne:2002st}, where the production of high-$p_T$ hadrons has been extensively studied experimentally \cite{PHENIX:2004nzn,PHENIX:2019gix,ALICE:2021est,PHOBOS:2004fsu,BRAHMS:2004xry,ATLAS:2016xpn,PHENIX:2017caf,LHCb:2021vww,LHCb:2022dmh,ALICE:2018vuu,STAR:2021fgw,CMS:2023snh,Braidot:2010zh,STAR:2006dgg,PHENIX:2011puq}. However, despite their theoretically simplicity, the phenomenological understanding of these processes remains challenging due to their reliance on non-perturbative fragmentation functions, and the interplay between initial and final state effects.

Over the past decade, high-energy jets and their substructure have become  powerful tools to probe  nuclear modifications \cite{Larkoski:2017jix,Asquith:2018igt,Marzani:2019hun,Cao:2020wlm,Apolinario:2022vzg, Cunqueiro:2021wls,Connors:2017ptx}. Jet substructure offers several phenomenological advantages, particularly due to its multi-scale nature, which provides  richer information than single-hadron measurements. A prominent example is energy correlators \cite{Basham:1979gh,Basham:1978zq,Basham:1978bw,Basham:1977iq}, a class of jet substructure observables defined as correlation functions of the average null energy (ANE) operator $\mathcal{E}(n)$ \cite{Sveshnikov:1995vi,Tkachov:1995kk,Korchemsky:1999kt,Bauer:2008dt,Hofman:2008ar,Belitsky:2013xxa,Belitsky:2013bja,Kravchuk:2018htv}. The one-point correlator measures the energy distribution as a function of direction, and simply corresponds to (moments of) inclusive single hadron production. Higher-point correlators provide access to transverse scales. The simplest of these, the two-point correlator (EEC), defines the scale, $p_T \theta~$ where $\theta$ is the angle between the two detectors and $p_T$  the initial hard scale, allowing the dynamics at specific physical scales to be isolated. These include the confinement scale \cite{CMS:2024mlf,Tamis:2023guc,ALICE:2024dfl,Komiske:2022enw}, the saturation scale \cite{Liu:2023aqb}, heavy-quark scales \cite{Craft:2022kdo,Andres:2023ymw,Xing:2024yrb}. Such sensitivity offers a much more detailed view into the interactions between energetic partons in hot and cold nuclear matter \cite{Andres:2022ovj,Andres:2023xwr,Andres:2023ymw,Yang:2023dwc,Barata:2023bhh,Andres:2024ksi,Bossi:2024qho,Andres:2024pyz,Andres:2024hdd,Xing:2024yrb,Singh:2024vwb,Devereaux:2023vjz,Fu:2024pic,Barata:2024wsu}. Energy correlators have recently been measured across various collider systems \cite{Komiske:2022enw,CMS:2024mlf,CMS-PAS-HIN-23-004,talkEEC,Tamis:2023guc,talk_Anjali,ALICE:2024dfl,talk_ALEPH}, revealing a clear nuclear modification in p-A and A-A collisions  relative to proton-proton (p-p) collisions \cite{CMS-PAS-HIN-23-004,talkEEC,Tamis:2023guc,talk_Anjali}.

Despite the significant phenomenological advantages of jet substructure observables, their theoretical understanding is complicated by the presence of transverse scales, in contrast to single-inclusive observables which are functions of longitudinal variables and can be cleanly expressed in terms of light-ray operators. Remarkably, energy correlators bridge this gap by giving access to transverse scales, such as $\theta$, while being formulated in terms of the ANE operator. This allows their transverse structure to be expanded using the light-ray OPE \cite{Hofman:2008ar}, reducing them to a series in $\theta$, with coefficients given by matrix elements of light-ray operators. 
Hence, this expansion establishes a precise correspondence between distinct scalings in $\theta$ and matrix elements of the underlying field theory. In the case of p-p collisions, the light-ray OPE has been highly successful, mapping the small-angle scaling in the perturbative regime of the energy correlators to the anomalous dimensions of twist-2 light-ray operators. This approach has enabled precision calculations \cite{Dixon:2019uzg,Lee:2022ige,Chen:2023zlx}, and the most precise extraction of $\alpha_s$ from jet substructure measurements \cite{CMS:2024mlf}. Motivated by this success, we are now led to apply the light-ray OPE to the more complex case of nuclear collisions.

In this \textit{Letter}, we take the first step in this direction by showing that the leading nuclear modification of the EEC in the perturbative regime is encoded in the light-ray OPE as enhanced matrix elements of twist-4 light-ray operators. This results in a predicted behavior for the ratio of the EEC in nuclear media (p-A or A-A) to vacuum of the from $\sim 1+a \theta^2$, where $a$ encodes properties of the nuclear medium. For jet radii $R$ used in experiments, this leading correction in the OPE is sufficient to describe the data in the range $\Lambda_{\rm QCD}/p_T < \theta \lesssim R$, as verified using CMS data from Pb-Pb collisions \cite{CMS-PAS-HIN-23-004} and ALICE data from p-Pb collisions \cite{talk_Anjali}. Our results highlight the potential of the light-ray OPE as a rigorous framework for understanding nuclear modifications of jet substructure observables in nuclear collisions.


\emph{The Light-Ray OPE.}---The light-ray OPE, introduced in \cite{Hofman:2008ar} and developed in conformal field theories (CFTs) in \cite{Kologlu:2019mfz,Chang:2020qpj}, has also been explored in QCD in \cite{Chen:2020adz,Chen:2022jhb,Chen:2021gdk,Chen:2023zzh}. Much like how the local OPE provides an expansion in the separation of local operators, the light-ray OPE provides an expansion in the angular separation $\theta$ of light-ray operators. This expansion is ideally suited for  jet substructure, where measurements are performed at small angles within the radius of an identified jet.

To use the light-ray OPE to understand nuclear collisions in real-world QCD,  it is essential to highlight key differences compared to its formulation in CFTs. Specifically, we must distinguish between differences in the structure of the OPE itself and modifications to the observable arising from performing the measurement in a hot or cold nuclear state.  The light-ray OPE, as an operator statement, is a property of the theory rather than the state. The dependence on the state is captured in the expectation values of the light-ray operators appearing in the OPE.  In CFTs, the light-ray OPE for two ANE operators takes the schematic form of an expansion \cite{Hofman:2008ar,Kologlu:2019mfz}
\begin{align}
\label{eq:OPE}
\mathcal{E}(n_1) \mathcal{E}(n_2)= \sum_{i} \theta^{\tau_i-4} \mathcal{C}_i  \,\mathbb{O}_i^{[3]}   \,,
\end{align}
with a scaling dictated by their twist $\tau_i$. The spin $J$ of the operator, fixed at $J=3$, is denoted in the superscript, \cite{Hofman:2008ar}.  The OPE coefficients $\mathcal{C}_i$  include kinematic twist effects, which are resummed into celestial blocks \cite{Kologlu:2019mfz,Chen:2022jhb,Chang:2022ryc}. In weakly-coupled gauge theories, the leading operators contributing to the angular scaling have classical twist 2 plus a quantum correction given by their anomalous dimension \cite{Hofman:2008ar}.  While we do not focus here on the explicit structure of the light-ray operators, recent progress in understanding their form, particularly beyond leading twist, is noteworthy \cite{Homrich:2024nwc,Ekhammar:2024neh}. 

In QCD, this OPE is modified both by the presence of the confinement scale $\Lambda_{\text{QCD}}$ and the non-vanishing $\beta$ function, both of which  lead to violations of the $J=3$ selection rule \cite{Dixon:2019uzg,Chen:2024nyc,Chen:2023zzh}.  Corrections of $\cO(\Lambda_{\text{QCD}})$ can be suppressed at high energies, while $\beta$-function corrections can be incorporated perturbatively. These effects are intrinsic to  the quantum field theory itself. To distinguish these intrinsic modifications from changes in the expectation values  of the light-ray operators, which reflect state-specific properties relevant in nuclear collisions, we use the light-ray OPE in its form \eqref{eq:OPE}. Modifications from hadronization and the running coupling can be included as needed but are omitted in this \textit{Letter} for clarity.

We emphasize that the state-independence of the OPE makes it a powerful tool for systematically understanding  the EEC in complex states, such as those produced in p-A and A-A collisions. By removing the transverse structure, the light-ray OPE 
maps contributions from specific light-ray operators to different power-law scalings. When  energy correlators are measured in a specific state $|\Psi \rangle$, the coefficients of these scalings in $\theta$ are determined by expectation values of the light-ray operators in that state
\begin{align}
\label{eq:OPE_state}
\langle \Psi |\mathcal{E}(n_1) \mathcal{E}(n_2) | \Psi \rangle = \sum_{i} \theta^{\tau_i-4} \mathcal{C}_i \langle \Psi |  \mathbb{O}_i^{[3]} | \Psi \rangle   \,.
\end{align}
This links the power-law behavior of the observable to well-defined matrix elements of light-ray operators, which can be computed to the desired order in the OPE expansion instead of computing the full observable directly.

\emph{Enhanced Higher Twist Effects in Nuclear Collisions.}---We now apply the light-ray OPE  to characterize the leading modifications of the EEC in nuclear collisions relative to its vacuum scaling. As the OPE of the EEC is fixed in \eqref{eq:OPE}, we see from \eqref{eq:OPE_state} that the nuclear modification is encoded in the relative size of the expectation values of light-ray operators with different twists. A useful analogue in $\mathcal{N}=4$ super-Yang-Mills helps us develop our intuition. In \cite{Chicherin:2023gxt}, the EEC was computed at weak coupling in heavy states created by the $1/2$ BPS operator $O_{\text{H}}(x)=\text{tr}[\phi^K(x)]$ with dimension $\Delta_{\text{H}}=K$. In the OPE limit, they found that
\begin{align}
\langle \mathcal{E}(n_1) \mathcal{E}(n_2) \rangle \sim \frac{ \langle \mathbb{O}_i^{[3]} \rangle_{\text{H}}   }{\theta^{2-\gamma(3)}} + 1 \,,
\end{align}
with $\langle \mathbb{O}_i^{[3]} \rangle_{\text{H}} \lesssim 1/\Delta_{\text{H}}$ for large $K$. This reveals the critical angle $\theta_*^{2-\gamma(3)}\sim 1/\Delta_{\text{H}}$,  determining the radius of convergence of the light-ray OPE. For $\theta< \theta_*$ we observe the characteristic twist-2 scaling, while for $\theta> \theta_*$, the uniform scaling characteristic of a twist-4 operator dominates. This clearly shows that the leading dependence on the state arises from the ratio between the twist-2 and twist-4 OPE coefficients.  Performed at weak coupling, this calculation can be viewed as perturbatively building a jet in a many particle background (where the operators twists are near their free values), suggesting that its behavior is in qualitative agreement with asymptotically free jets modified by a nuclear state.

We now apply this intuition to understand the modification of energy correlators in nuclear collisions.  Unlike the case of heavy operators in $\mathcal{N}=4$, where the expectation values of light-ray operators and the radius of convergence of the light-ray OPE are set by the single parameter $\Delta_H$, nuclear collisions involve multiple dimensionful parameters characterizing the state. These include transport coefficients $\hat q$, the temperature $T$, the size of the QGP (nucleus), $L$ ($A$), etc. Nevertheless,  a critical angle must exist below which the OPE converges, with leading nuclear effects  encoded in the ratio of expectation values of the twist-2 and twist-4 operators. Multiple characteristic angles, such as
$\theta_{L}\sim 1/\sqrt{p_T L}$ and $\theta_{c}\sim 1/\sqrt{\hat{q} L^3}$, have appeared in EEC calculations \cite{Mehtar-Tani:2010ebp,Casalderrey-Solana:2011ule,Mehtar-Tani:2012mfa,Casalderrey-Solana:2012evi,Mehtar-Tani:2017ypq, Caucal:2018dla,Andres:2022ovj,Andres:2023xwr,Andres:2023ymw,Barata:2023bhh}, and  the light-ray OPE is expected to converge for $\theta$ small enough so the medium scales do not dominate the EEC entirely. Nuclear effects are well known to enhance higher-twist contributions in inclusive hadron production \cite{Luo:1993ui,Kastella:1989ux,Luo:1992fz}, with terms in the twist expansion enhanced by powers of $A^{1/3}$ in cold nuclear matter. This  closely aligns with the intuition from $\mathcal{N}=4$.

To characterize the leading nuclear effects in the EEC, we first make the approximation, well justified from analytic calculations and data \cite{Tamis:2023guc,ALICE:2024dfl,CMS:2024mlf,Lee:2022ige,Chen:2023zlx}, that the small-angle perturbative ($ \Lambda_{\rm QCD}/p_T<\theta < 1$) scaling of the EEC in vacuum is governed by  leading twist operators  
\begin{align}
\langle \Psi | \mathcal E(n_1)  \mathcal E(n_2) | \Psi \rangle \sim \frac{1}{\theta^2} \langle \Psi |\vec{\mathcal{C}}_{2} \cdot \vec{\mathbb{O}}_{\tau=2}^{[3]} | \Psi \rangle  \,.
\label{EEC_vac}
\end{align}
Modifications of the EEC scaling behavior in p-A and A-A collisions must then arise from enhanced contributions from higher-twist operators. In the perturbative regime,  twist-3 corrections are absent \cite{Jaffe:1991ra,Dixon:2019uzg,Chen:2022jhb}, so the leading corrections appear at twist-4
\begin{align}
 \langle \Psi_{\rm N} | \mathcal E(n_1)  \mathcal E(n_2)| &\Psi_{\rm N} \rangle \sim   
 \langle \Psi_{\rm N} |  \frac{\vec{\mathcal{C}}_{2} \cdot \vec{\mathbb{O}}_{\tau=2}^{[3]}}{\theta^2}  +   \vec{\mathcal{C}}_{4} \cdot \vec{\mathbb{O}}_{\tau=4}^{[3]} | \Psi_{\rm N} \rangle\,.
 \label{EEC_med}
\end{align}
Here, $\Psi_N$ denotes a state produced in a nuclear collision. In general, there are multiple twist-4 operators. Since we do not study their renormalization in this paper, they are all degenerate, and we can view this as their total matrix element. Taking the ratio of the EEC in the nuclear medium  \eqref{EEC_med} to that in vacuum \eqref{EEC_vac}, yields a simple prediction for the leading deviations due to nuclear effects
\begin{align}
 &\frac{  \langle \Psi_{\rm N} | \mathcal E(n_1)  \mathcal E(n_2)| \Psi_{\rm N} \rangle }{ \langle \Psi | \mathcal E(n_1)  \mathcal E(n_2)| \Psi \rangle} \label{eq:OPEfit} \\
  &\hspace{0.5cm} \sim \frac{ \langle \Psi_{\rm N} | \vec{\mathcal{C}}_{2} \cdot \vec{\mathbb{O}}_{\tau=2}^{[3]}  | \Psi_{\rm N} \rangle   }{  \langle \Psi |  \vec{\mathcal{C}}_{2} \cdot \vec  {\mathbb{O}}_{\tau=2}^{[3]}    | \Psi \rangle    }  +\theta^2   \frac{ \langle \Psi_{\rm N} | \vec{\mathcal{C}}_{4} \cdot \vec{\mathbb{O}}_{\tau=4}^{[3]} | \Psi_{\rm N} \rangle    }{  \langle \Psi |    \vec{\mathcal{C}}_{2} \cdot \vec{\mathbb{O}}_{\tau=2}^{[3]}    | \Psi \rangle    }  \,. \nonumber 
\end{align}
The remarkable aspect of the light-ray OPE is its prediction of the functional form of the leading nuclear modifications, which are quadratic in $\theta$, with the coefficient given by the expectation value of a twist-4 operator in the nuclear state. This matrix element can either be fitted to data or analytically computed, enabling the determination of its dependence on the state's properties. We will show that, for jet radii used in experiments, including  up to the twist-4 contribution provides an excellent description of the data.

\emph{Direct applications to measurements: A-A.}---We now apply the light-ray OPE to understand the $\theta$-dependence of the recent   EEC measurement in Pb-Pb collisions \cite{talkEEC,CMS-PAS-HIN-23-004}, which reveals a clear nuclear enhancement. This demonstrates the OPE's power: even in  complex nuclear collisions, nuclear modifications follow the scaling in \eqref{eq:OPEfit}. 

Our analysis examines the modifications of the EEC in Pb-Pb relative to p-p using the OPE for each individual spectrum, a method that significantly reduces experimental systematics. For this approach to be valid, both expansions must share the same domain of convergence. However, experimental selection-bias shifts the non-perturbative transition in the Pb-Pb EEC, misaligning the lower bound of the perturbative window in $\theta$ between the two samples. To maximize the extent of the perturbative window in the Pb-Pb/p-p ratio, minimizing the effects of selection bias is thus essential. To this end, we follow the procedure in \cite{Andres:2024hdd, Andres:2024pyz}, correcting the Pb-Pb EEC spectrum with the factor $C_2$, which is classically one in the small-angle perturbative regime. Should higher $p_{T}$ measurements become available, where the misalignment diminishes, the need for this correction will decrease.

\begin{figure}
\includegraphics[width=0.45\textwidth]{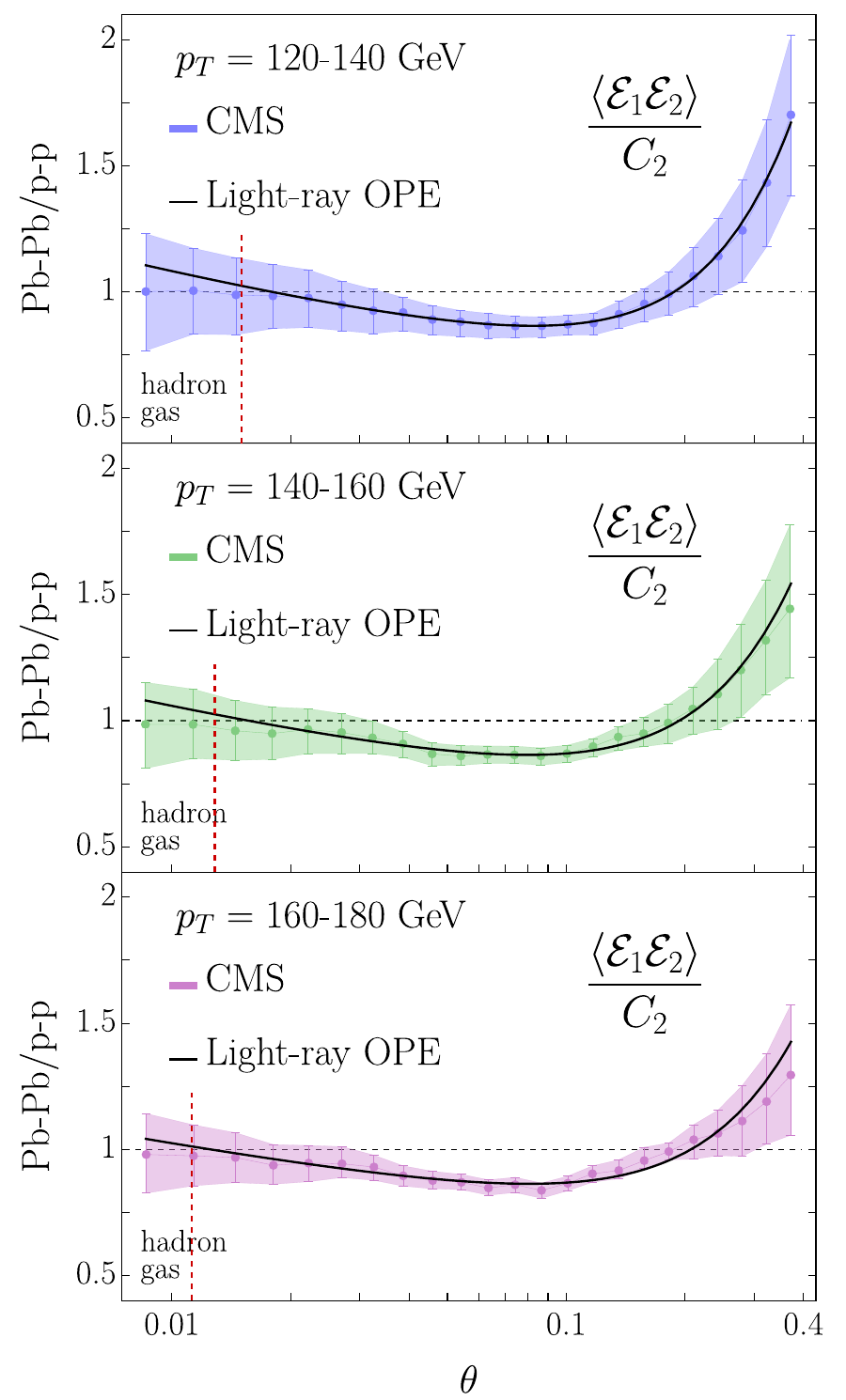}
\caption{Top: Ratio of the EEC/$C_2$ for inclusive jets with
reconstructed $p_T$ in the range $120 < p_T < 140~$GeV within the 0-10$\%$ centrality class in  $\sqrt{s_{\rm NN}} = 5.02$ TeV Pb-Pb collisions over that of p-p jets (colored band) compared to the OPE fit (black). The EEC/$C_2$ distribution is obtained from the CMS EEC measurement \cite{talkEEC,CMS-PAS-HIN-23-004} using the \textit{unbiasing} procedure from \cite{Andres:2024hdd, Andres:2024pyz}. Middle (Bottom): same as  top for $140<p_T < 160$ GeV ($160<p_T < 180$ GeV). }
\label{fig:AA}
\end{figure}

Since the expectation values of the twist-4 operators $ \langle \Psi_{\rm N} | \mathbb{O}_{\tau=4}^{[J=3]} | \Psi_{\rm N} \rangle$ in different states are  unknown, we treat them as parameters to be fitted to the data in the lowest jet $p_{T}$ bin. A common feature of twist expansions is that matrix elements are accompanied by factors of $p_{T}^{-\tau}$ in the large-$p_{T}$ limit, fixed classically by dimensional analysis \cite{Luo:1993ui,Luo:1994np,Luo:1992eq,Qiu:1991wg,Luo:1992fz,Luo:1991bj,Kastella:1989vd,Kastella:1989ux}. We thus determine the twist-4 matrix elements in higher $p_{T}$ bins by scaling the fitted result from the lowest bin according to this power law. To describe the data, we must also incorporate the anomalous dimensions of the twist-two operators arising from their renormalization \cite{Dixon:2019uzg}. Twist-4 operators are treated as classical, omitting extra parameters for their presently unknown anomalous dimensions. Including these dimensions will be an interesting future direction, motivating further exploration of their structure \cite{Braun:2001qx,Braun:2000yi,Derkachov:1999ze,Braun:2000av,Braun:1999te,Homrich:2024nwc,Ekhammar:2024neh,Belitsky:1999bf,Belitsky:1999qh}.

In \Fig{fig:AA}, we show the Pb-Pb/p-p EEC ratio corrected by the \textit{unbiasing} function $C_2$ \cite{Andres:2024hdd, Andres:2024pyz} compared to the power-law scaling in \eqref{eq:OPEfit}, effectively of the form $\sim 1+a\theta^2$. The details of our fitting procedure are provided in the Supplemental Material, along with results for other $p_{T}$ and centrality bins. We see that the combination of the first two leading terms in the OPE (black curve) describes the data very well, with no free parameters required to describe the wide-angle enhancement in the middle and bottom panels of \Fig{fig:AA}. It is genuinely remarkable that such a simple power-law can be identified in these complex collisions, as it also allows the measurement of the effective twist-4 matrix element characterizing it, which is sensitive to the properties of the medium. Additionally, this scaling reduces the calculation of the EEC to determining this matrix element, rather than determining a full function of $\theta$.


\emph{p-A collisions.}---The first measurement of  energy correlators in p-A has recently been reported \cite{talk_Anjali}, showing an unexpectedly large modification  with respect to p-p in the lowest $p_T$ bin. As in Pb-Pb, this measurement can also be understood using the OPE in \eqref{eq:OPEfit}, with the $p_T$ dependence of the modification predicted by the twist of the light-ray operators.

This measurement  highlights another benefit of  a first-principles theoretical understanding of nuclear modifications being a power law.  In addition to power-law modifications, one also expects \emph{logarithmic} modifications due to changes in the quark/gluon fractions in p-A relative to p-p, which we include as detailed in the Supplemental Material. In \Fig{fig:pA} we show fits to the data compared with the OPE \eqref{eq:OPEfit} (black), as well as the twist-2 leading logarithmic prediction with a modified quark/gluon fraction (green). We note that in the middle and bottom panels, these curves coincide. We emphasize the importance of including both  the anomalous dimensions and the twist-4 term in this fit in the lowest $p_{T}$ bin. Using leading-log resummation \cite{Dixon:2019uzg}, one can solve for the extracted quark/gluon fractions in the OPE fit. In the Supplemental Material, we demonstrate that by assuming an even q/g fraction in p-p, one finds an approximate p-A q/g fraction of $30/70$ consistent with expectations from nuclear PDFs \cite{AbdulKhalek:2022fyi,Eskola:2021nhw}. The importance of the anomalous dimensions highlights that a proper theoretical understanding will be crucial for fully understanding measurements of energy correlators in p-A, as well as in other small systems. We believe that the p-A results deserve a more detailed study, beyond the scope of this \emph{Letter}.

\begin{figure}
\includegraphics[width=0.45\textwidth]{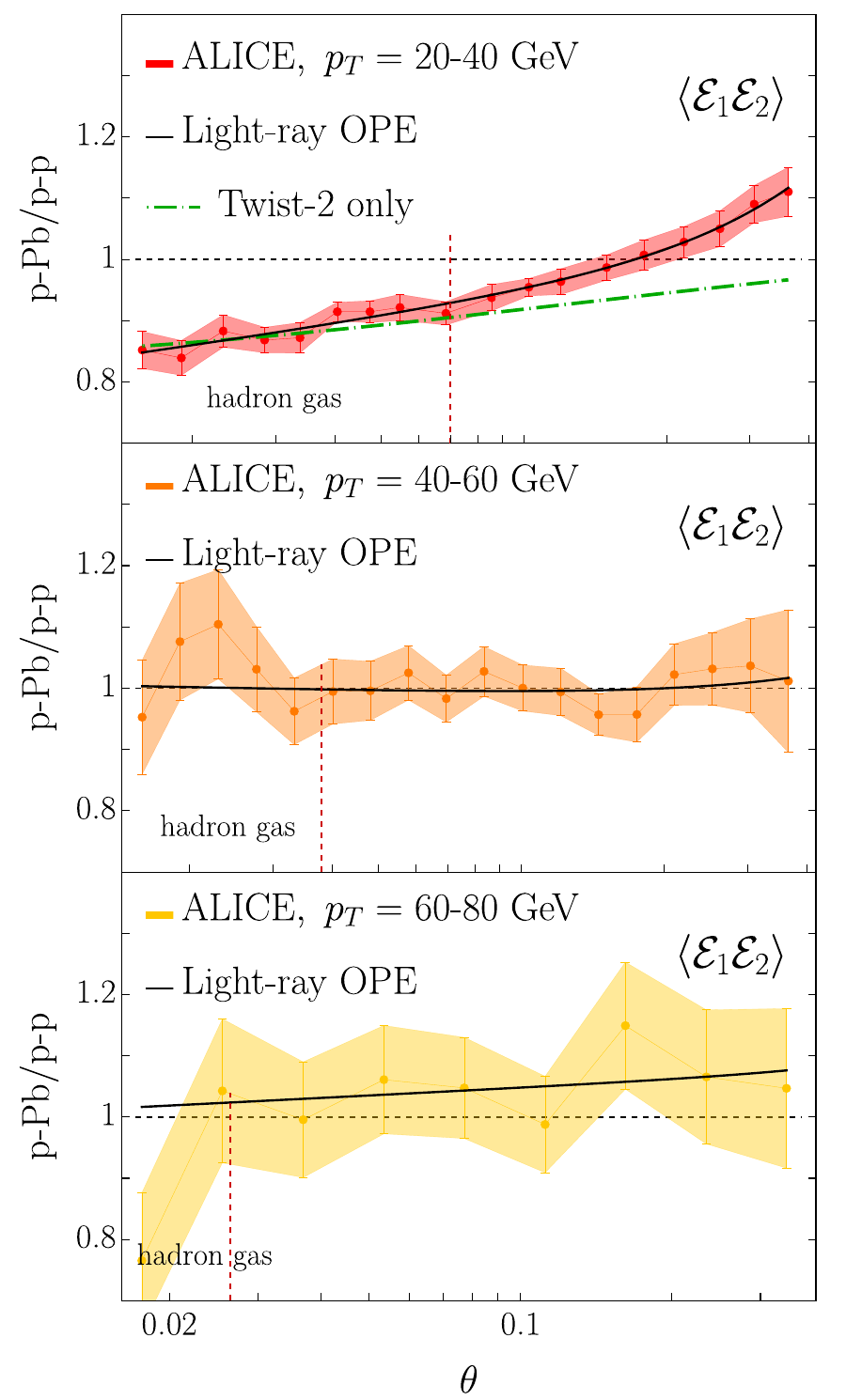}
  \caption{Top: Ratio of the EEC for inclusive jets with reconstructed $p_T$ in the range $20<p_T<40$ GeV in  $\sqrt{s_{\rm NN}} = 5.02$ TeV p-Pb collisions over that of p-p jets (colored band) compared to the OPE (twist-2+ twist-4) fit (black solid) and to the twist-2  contribution only (green dot-dashed). Middle (Bottom): same as  top for $40<p_T < 60$ GeV ($60<p_T < 80$ GeV).}
  \label{fig:pA}
\end{figure}

\emph{Factorizing the Light-Ray Density Matrix.}---Having highlighted how the light-ray OPE allows a systematic expansion around the small angle limit, predicting definite scaling laws with coefficients given by expectation values of light-ray operators, we now want to show how the calculation of these matrix elements relates to higher twist inclusive hadron production. We believe that this is an important step in relating energy correlators to the highly developed factorization theorems for inclusive hadron production, and will facilitate the calculation of these matrix elements within specific models.

\begin{figure}
\includegraphics[width=0.45\textwidth]{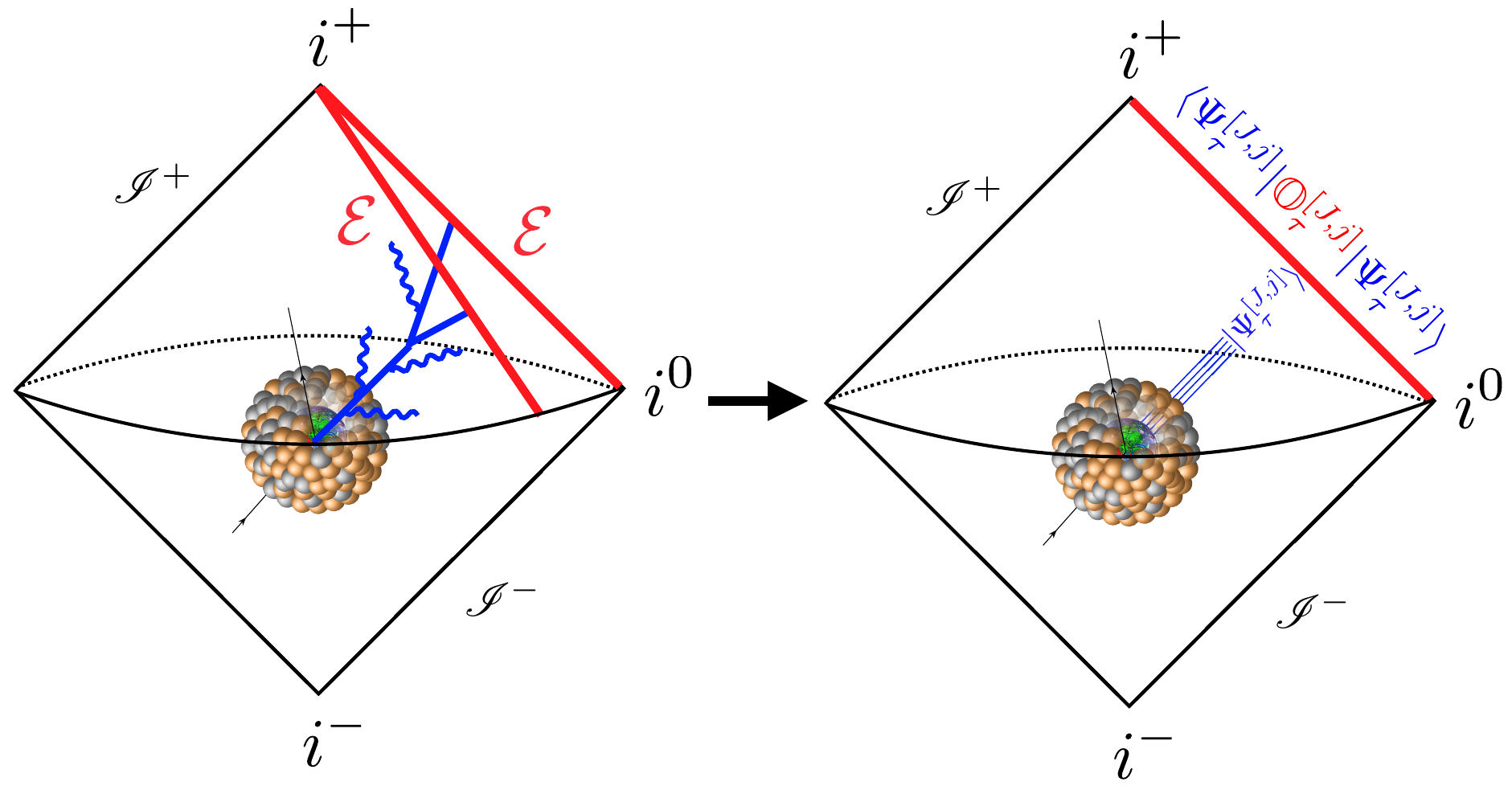}
  \caption{The OPE expresses the energy correlator as a sum over expectation values of light-ray operators. These can be computed in nuclear collisions using factorization theorems for the dual light-ray densities, which are projections of higher-twist fragmentation onto specific quantum numbers. }
  \label{fig:QGP_OPE}
\end{figure}

After applying the light-ray OPE, we must compute expectation values of specific light-ray operators in a given state, which so far we have left generic. The difficulty with nuclear collisions, is that this state is produced by the collision of nuclei, instead of a local operator, and therefore must be computed using factorization theorems. Taking for concreteness the collision of two nuclei, $A$, with momenta $p_1,p_2$, we would like to compute the following expectation value
\begin{align}
\langle A(p_1) A(p_2)| \mathbb{O}_{\tau}^{[3]} | A(p_1) A(p_2) \rangle\,.
\end{align}
Here another simplification occurs, due to the fact that the light-ray operator has definite quantum numbers under the Lorentz group, particularly the celestial quantum numbers, namely the spin, $J$, associated with longitudinal boosts, and the transverse spin, $j$ associated with transverse rotations. Since we focus here only on unpolarized measurements, we project onto $j=0$, and drop this quantum number for simplicity. We can therefore insert a complete set of states dual to the light-ray operators, which are referred to as light-ray densities \cite{Chang:2022ryc}.  We then have
\begin{align}
\hspace{-0.32cm}\sum_{J, J'} \langle A(p_1) A(p_2)| \Psi_{\tau}^{[J]} \rangle \langle \Psi_{\tau}^{[J]} | \mathbb{O}_{\tau}^{[3]} | \Psi_{\tau}^{[J']} \rangle   \langle \Psi_{\tau}^{[J']}   | A(p_1) A(p_2) \rangle.
\end{align}
Since the states $|\Psi_{\tau}^{[J]} \rangle$ are eigenvectors of the operator $\mathbb{O}_\tau^{[J]}$, the measurement of $\mathbb{O}_\tau^{[3]}$  projects onto the single state $|\Psi_{\tau}^{[3]} \rangle$. Therefore, we only have to compute
\begin{align}
|\langle A(p_1) A(p_2)| \Psi_{\tau}^{[3]} \rangle |^2\,.
\end{align}
 However, this is simply a (higher twist) inclusive hard function projected onto specific quantum numbers. At twist-2 the light-ray densities are single particle states, relating the calculation to inclusive hadron production at leading twist, as used in \cite{Dixon:2019uzg,Lee:2022ige}. At twist-4 they are up to two-particle states. This factorization is illustrated in \Fig{fig:QGP_OPE}.

The light-ray OPE has therefore reduced the calculation of multi-point energy correlators at the LHC, to computing the probability to produce specific multi-particle states $|\Psi_{\tau}^{[3]} \rangle$. As mentioned in the introduction, these are the objects that are under the best theoretical control. Particularly in e-A and p-A they are described by higher twist factorization theorems
\cite{Politzer:1980me,Ellis:1982cd,Ellis:1982wd,Jaffe:1983hp,Jaffe:1981td,Jaffe:1982pm,Qiu:1990xy,Qiu:1990xxa} whose matrix elements are indeed enhanced by the nuclear size, $A^{1/3}$ \cite{Luo:1993ui,Luo:1994np,Luo:1992eq,Qiu:1991wg,Luo:1992fz,Luo:1991bj,Kastella:1989vd,Kastella:1989ux}. So far these calculations have been performed with approximate hard functions that are derivatives of the four-parton hard function \cite{Kang:2013ufa,Luo:1992fz,Luo:1991bj,Luo:1994np}. It will be interesting to extend this to the full calculation. Such calculations can also be performed within the color glass condensate (CGC) framework \cite{Iancu:2003xm,McLerran:1993ni,McLerran:1993ka,McLerran:1994vd}.
Strictly speaking, if the state is identified using a jet algorithm, then this factorization must be done for hadron production inside an identified jet using the fragmenting jet formalism \cite{Kang:2016ehg,Kang:2016mcy,vanBeekveld:2024jnx,vanBeekveld:2024qxs,Liu:2022ijp,Lee:2024icn,Lee:2024tzc}. We find it particularly exciting that the light-ray OPE allows us to relate the calculation of jet substructure observables to these well understood objects, putting the calculation of energy correlators in nuclear collisions on firm theoretical footing.  We will pursue a full calculation of these matrix elements in different models in a future paper.

\emph{Conclusions.}---Energy correlator observables have shown immense potential for understanding the dynamics of nuclear collisions, as their dependence on $\theta$ allows nuclear systems to be probed as a function of scale. However, the inherent complexity of nuclear collisions poses significant challenges to the calculation of the full $\theta$ dependence of these observables from first-principles theory.

Energy correlators can be formulated directly in terms of light-ray operators, in contrast to traditional jet substructure observables. In this \emph{Letter} we have shown that the light-ray OPE provides an organizing principle for understanding the modification of energy correlators in nuclear collisions, by enabling a systematic expansion in the angle $\theta$, with the coefficients of distinct power-law scalings associated with matrix elements of light-ray operators.  We showed that the leading nuclear modification is associated with enhanced twist-4 light-ray operators, and that by incorporating just these operators, we are able to predict the $\theta$ dependence within the range of angles relevant for jet substructure studies. Our prediction beautifully describes both the recent CMS data for the EEC in A-A collisions, and the ALICE data for the EEC in p-A.

From the theoretical perspective, our analysis motivates the exploration of the light-ray OPE in other states, such as thermal, large charge \cite{Firat:2023lbp}, or plasma ball states  \cite{Aharony:2005bm}.
From the experimental perspective, our analysis motivates the experimental measurement of higher point correlators, both shape dependent and projected, in nuclear collisions. These were studied in the context of the hybrid model in \cite{Bossi:2024qho}, but can now be brought under analytic control using the light-ray OPE approach. In particular, ratios of projected correlators \cite{Chen:2020vvp} will access the anomalous dimensions of twist-4 light-ray operators, enabling these twist-4 light-ray operators to be studied in high-quality data in the perturbative regime. It would also be interesting to study different detectors involving higher energy weights, or charges \cite{Lee:2023tkr,Lee:2023npz}. The light-ray OPE approach introduced in this \emph{Letter} allows us to theoretically understand this broader class of observables, which we hope will improve our understanding of exotic states of nuclear matter.

\emph{Acknowledgments.}--- 
 We would like to thank Hannah Bossi,  Hao Chen, Gabriel Cuomo, Wenqing Fan, Laura Havener, Barbara Jacak,  Murat Kologlu, Juan Maldacena,  Isaac Mooney, Anjali Nambrath,  Mateusz Ploskon, Ananya Rai, Krishna Rajagopal,  Marta Verweij, Xin-Nian Wang and Sasha Zhiboedov for useful discussions. This work is supported in part by the GLUODYNAMICS project funded by the ``P2IO LabEx (ANR-10-LABX-0038)'' in the framework ``Investissements d'Avenir'' (ANR-11-IDEX-0003-01) managed by the Agence Nationale de la Recherche (ANR), France. The work of CA was partially supported by the U.S. Department of Energy, Office of Science, Office of Nuclear Physics under grant Contract Number DE-SC0011090. FD is supported by the European Research Council under project ERC-2018-ADG-835105 YoctoLHC; by Maria de Maeztu excellence unit grant CEX2023-001318-M and project PID2020-119632GB-I00 funded by MICIU/AEI/10.13039/501100011033; and by ERDF/EU. FD has received funding from Xunta de Galicia (CIGUS Network of Research Centres). JH is supported by the Leverhulme Trust as an Early Career Fellow. 

\bibliography{EEC_ref.bib}{}
\bibliographystyle{apsrev4-1}

\begin{widetext}

\section*{Supplemental Material}
In this \emph{Supplemental Material}, we provide a more detailed discussion of the fits of the A-A/p-p and p-A/p-p  energy correlator ratios. Prior to renormalization, the OPE takes the form
\begin{align}
 &\frac{ \langle \Psi_{\rm N} | \mathcal E(n_1)  \mathcal E(n_2)| \Psi_{\rm N} \rangle }{ \langle \Psi | \mathcal E(n_1)  \mathcal E(n_2)| \Psi \rangle} \sim  \frac{ \langle \Psi_{\rm N} | \vec{\mathcal{C}}_{2} \cdot \vec{\mathbb{O}}_{\tau=2}^{[3]}  | \Psi_{\rm N} \rangle}{  \langle \Psi |  \vec{\mathcal{C}}_{2} \cdot \vec  {\mathbb{O}}_{\tau=2}^{[3]}    | \Psi \rangle    }  + \theta^2   \frac{ \langle \Psi_{\rm N} | \vec{\mathcal{C}}_{4} \cdot \vec{\mathbb{O}}_{\tau=4}^{[3]} | \Psi_{\rm N} \rangle}{\langle \Psi |    \vec{\mathcal{C}}_{2} \cdot \vec{\mathbb{O}}_{\tau=2}^{[3]}    | \Psi \rangle} \,, 
 \label{eq:OPE_clas}
\end{align}
where  the matrix elements of the light-ray operators for a given twist are degenerate. Renormalizing the light-ray operators breaks this degeneracy by introducing the anomalous dimensions $\gamma_{ij}$,
\begin{align}
    \mathbb{O}^{[J]}_{i, \,{\rm bare}}= \mathbb{Z}_{ij} \mathbb{O}^{[J]}_{j, \,{\rm renormalized}}\,,
    \qqquad
    \frac{\td \, \mathbb{O}^{[J]}_{i, ~{\rm renormalized}}}{\td \ln \mu^2} = \gamma_{ij} \mathbb{O}^{[J]}_{j, \,{\rm renormalized}} \,,
\end{align}
where, in our case, $\mu = \theta p_{T}$. This renormalization breaks the degeneracy between quark-like and gluonic operators, causing the scaling laws to be  modified through the anomalous dimensions by the effective quark/gluon fraction in the process. This effect becomes important when comparing measurements across different processes (A-A versus p-p) or energies, which can involve different quark/gluon fractions. Consequently, we must modify the classical OPE  \eqref{eq:OPE_clas} to include these effects. For clarity, we assume  fixed coupling, as including the QCD $\beta$-function is a small modification that is well understood \cite{Dixon:2019uzg,Andres:2023xwr}. Renormalizing the twist-2 operators modifies \eqref{eq:OPE_clas} as
\begin{align}
 &\frac{  \langle \Psi_{\rm N} | \mathcal E(n_1)  \mathcal E(n_2)| \Psi_{\rm N} \rangle }{ \langle \Psi | \mathcal E(n_1)  \mathcal E(n_2)| \Psi \rangle} \sim \frac{ \langle \Psi_{\rm N} | \vec{\mathcal{C}}_{2} \cdot \theta^{\hat{\gamma}(3)} \cdot \vec{\mathbb{O}}_{\tau=2}^{[3]}  | \Psi_{\rm N} \rangle   }{  \langle \Psi |  \vec{\mathcal{C}}_{2} \cdot \theta^{\hat{\gamma}(3)} \cdot \vec  {\mathbb{O}}_{\tau=2}^{[3]}    | \Psi \rangle}  +\theta^2   \frac{ \langle \Psi_{\rm N} | \vec{\mathcal{C}}_{4} \cdot \vec{\mathbb{O}}_{\tau=4}^{[3]} | \Psi_{\rm N} \rangle    }{  \langle \Psi |    \vec{\mathcal{C}}_{2} \cdot \theta^{\hat{\gamma}(3)} \cdot \vec{\mathbb{O}}_{\tau=2}^{[3]}    | \Psi \rangle} \,,
 \label{eq:OPE_quant}
\end{align}
where $\hat{\gamma}(3)$ is the twist-2 spin-3 anomalous dimension matrix. We treat the twist-4 operator as classical, since its effect is a small perturbation to the leading twist-2 term.

As previously stressed, the anomalous dimensions themselves do not depend on the system studied. However, the relative dominance of quark-like operators versus gluonic operators will depend on the state in which they are evaluated. It is important to note that only the total sum of these matrix elements, along with their anomalous scaling which enters into the OPE \eqref{eq:OPE_quant}. Therefore, to simplify the OPE before fitting, we introduce an effective anomalous dimension  that captures the effect of both quark and gluon fractions:
\begin{align}
    \left[\langle \Psi_{ i} | \vec{\mathcal{C}}_{2} \cdot \theta^{\hat{\gamma}(3)} \cdot \vec{\mathbb{O}}_{\tau=2}^{[3]}  | \Psi_{ i} \rangle \right]_{\rm gluonic} + \left[\langle \Psi_{ i} | \vec{\mathcal{C}}_{2} \cdot \theta^{\hat{\gamma}(3)} \cdot \vec{\mathbb{O}}_{\tau=2}^{[3]}  | \Psi_{ i} \rangle \right]_{\rm quark-like} \approx M^{i}_{2} ~ \theta^{\gamma_{i}} \,.
    \label{eq:11} 
\end{align}
Thus, the basic functional form we use to fit OPE is
\begin{align}
    {\rm N/pp} = \frac{\mathcal{M}_2 ~ \theta^{\gamma_N} + \mathcal{M}_4 ~\theta^2}{\theta^\gamma}\,,
    \label{eq:}
\end{align}
where $N$ represents either the A-A or p-A process, and we have defined $\mathcal{M}_2 \equiv M^N_2/M_2$  and $\mathcal{M}_4 \equiv \langle \Psi_{\rm N} | \vec{\mathcal{C}}_{4} \cdot \vec{\mathbb{O}}_{\tau=4}^{[3]} | \Psi_{\rm N} \rangle / M_2$, with $M_2$ referring to p-p. A priori, this expression has four free parameters. $\mathcal{M}_4$ is initially unknown, but we know from the twist of the matrix elements how it scales with the jet $p_{T}$. $\mathcal{M}_2$, $\gamma_N$ and $\gamma$ are also unknown in general, as their calculation depends on the description of the state in which the OPE is evaluated. However, their classical values are known to be $1$, $0$, and $0$, respectively. In the perturbative region, derivations from these classical values are expected to be small. Therefore, we  constrain the fits such that $\mathcal{M}_2\in[0.5,1.5]$, $\gamma_N \in [-0.5,0.5]$, and $\gamma \in [-0.5,0.5]$. 

In summary, we fit both the A-A/p-p and p-A/p-p data with the following ansatz
\begin{align}
    {\rm N/pp} =  \frac{\mathcal{M}_2 ~ \theta^{\gamma_N} + \left(\frac{ p_{T,0}}{p_{T}}\right)^2 \mathcal{M}_4 (p_{T,0}) ~ \theta^2}{\theta^\gamma}\,,
    \label{eq:power-lawAnsatz}
\end{align}
where we have made explicit the $p_T$-dependence of the twist-4 matrix element relative to the twist-2 term. Here, $\mathcal{M}_4 (p_{T,0})$ is fitted to a particular reference bin in $p_T$, $p_{T,0}$,  for p-A/p-p  and to a particular reference bin in $p_T$, $p_{T,0}$,  for each centrality class in A-A/p-p data. The reference value of  $p_{T,0}$ is chosen to be the lowest $p_T$ bin for which data are available, both for the p-A/p-p and A-A/p-p fits. Consequently, in p-A, $\mathcal{M}_4 (p_{T,0})$ remains constant across all $p_T$ bins. This means that the   modification (if any) obtained in the two higher $p_T$ bins in p-A, relative to p-p, is a predicted outcome of the $p_T$-dependence of the twist of the matrix elements.  Similarly, in the A-A/p-p fits, $\mathcal{M}_4$ remains constant across all $p_T$ bins within a given centrality class. As a result, the size of the A-A enhancement relative to p-p is a prediction for all $p_T$ bins within each centrality class, except for the bin used to determine $\mathcal{M}_4 $. Additionally, since $\gamma$ is determined by the p-p baseline and should therefore be independent of centrality, it is allowed to vary only as a function of $p_T$. In the A-A/p-p fits, $\gamma$ is thus fixed for each $p_T$ bin across all centralities to the value obtained from the most central class. In the future, this analysis could be improved by directly computing the q/g fractions and twist-4 anomalous dimensions. With these additional inputs, the only remaining parameter would be $\mathcal{M}_4 (p_{T,0})$, which is strongly dependent on the properties of the process. As such, $\mathcal{M}_4 (p_{T,0})$ would  provide a highly sensitive probe of the nuclear medium.

\begin{figure}
\includegraphics[width=0.99
\textwidth]{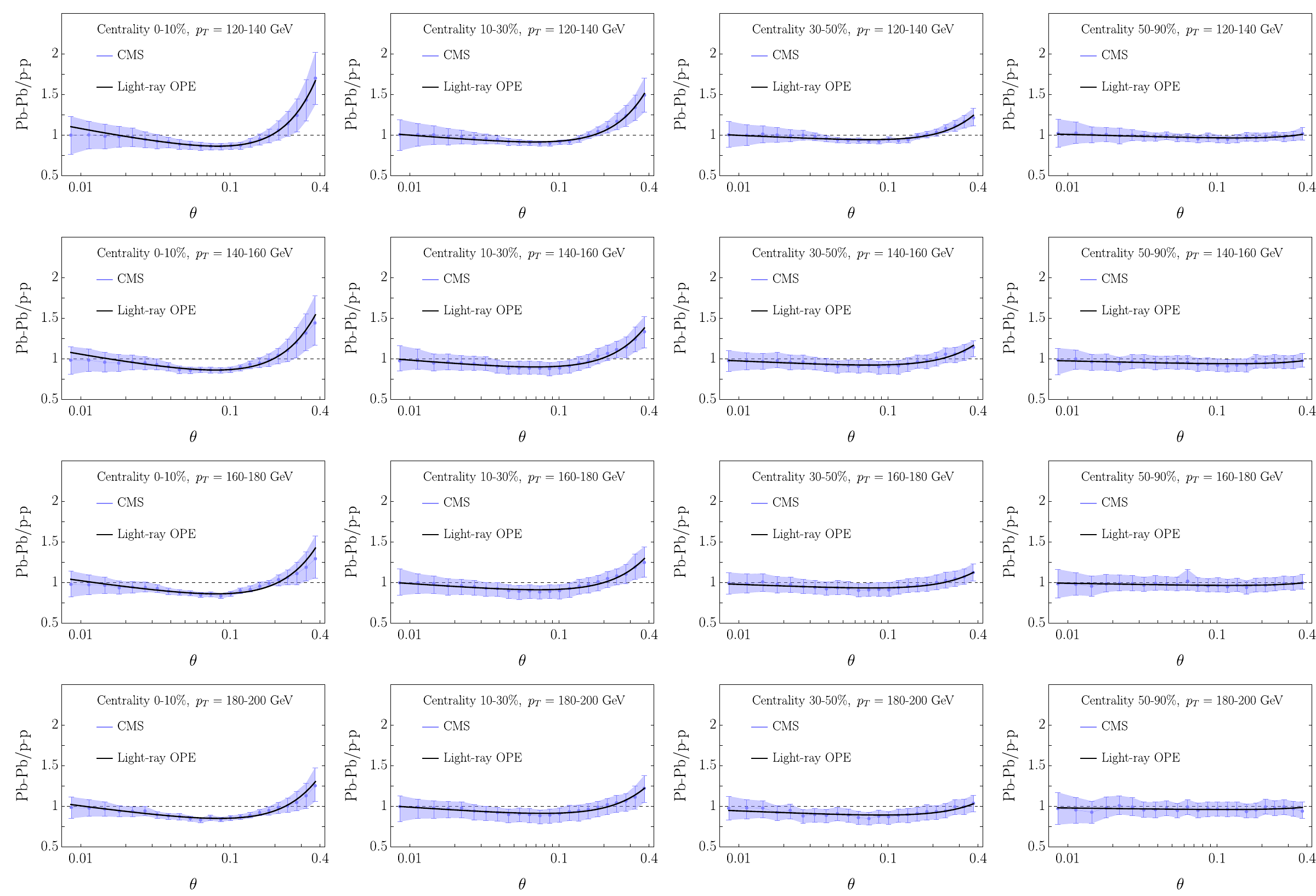}
\caption{Ratio of the EEC/$C_2$ for inclusive jets with
reconstructed $p_T$ in the range $120 < p_T < 200~$GeV with bins of $20~$GeV within the 0-10$\%$, 10-30$\%$, 30-50$\%$, 50-90$\%$ centrality classes in  $\sqrt{s_{\rm NN}} = 5.02$ TeV Pb-Pb collisions over that of p-p jets (blue band) compared to the OPE fit (black curve). The EEC/$C_2$ distribution is obtained from the CMS EEC measurement \cite{talkEEC,CMS-PAS-HIN-23-004} using the \textit{unbiasing} procedure from \cite{Andres:2024hdd, Andres:2024pyz}.}
\label{fig:AA_sup}
\end{figure}

In Fig.~\ref{fig:AA_sup} we show the OPE fitted to the \emph{unbiased} CMS Pb-Pb/p-p EEC measurement \cite{CMS-PAS-HIN-23-004,Andres:2024hdd,Andres:2024pyz} across 4 bins in $p_T$ and 4 bins in centrality. We use the 120-140~GeV $p_T$ bin (the top row) to find $\mathcal{M}_4 (120\mbox{-}140\,{\rm GeV})$ for each centrality class. The functional form of the wide angle enhancement is then entirely fixed in rows 2-4. It is clear that excellent agreement with the experimental data is found. Comparing to the analytic calculation with twist-2 leading-log resummation (which will be discussed in the following paragraph) we found that anomalous dimensions are consistent with a fractionally higher quark fraction in Pb-Pb than in p-p of $\sim 10\%$. We note that setting $\mathcal{M}_4(120\mbox{-}140\,{\rm GeV}) = 0$ fails to describe the data, indicating that twist-4 contributions are essential.

In Fig.~\ref{fig:pA} of the main body of the \emph{Letter} we apply the OPE ansatz to the p-A/p-p ratio. As was discussed in the main body of this \emph{Letter}, in this case the effective anomalous dimensions ($\gamma_{pA}$ and $\gamma$) play a large roll in the lowest $p_{T}$ bin. Using leading resummation for the left-hand-side of \eqref{eq:11}, one can solve for the quark fraction in p-A in terms of the p-p quark fraction which results in the measured values of $\gamma_{pA}$ and $\gamma$. The relevant twist-2 leading-log formulae are derived in full in the appendices of \cite{Andres:2023xwr}. The relevant extracted fit parameters were $\gamma_{pA} =0.34$ and $\gamma=0.29$  are consistent with $N^{pA}_q \approx -0.05 + 0.85 N^{pp}_q$, where we normalize the q/g fractions so that $N_q+N_g=1$. Therefore, assuming an even q/g fraction in p-p, one finds an approximate p-Pb q/g fraction of $30/70$ consistent with expectations from nuclear PDFs \cite{AbdulKhalek:2022fyi}. An important observation in the p-Pb/p-p fit is that the $p_{T}$-suppression of the twist-4 contribution has a dominant effect in the higher $p_{T}$ bins. This justifies the absence of any enhancement in p-Pb/p-p ratio for higher jet $p_{T}$.

\end{widetext}


\end{document}